\documentclass[sigconf]{acmart}

\AtBeginDocument{%
  \providecommand\BibTeX{{%
    \normalfont B\kern-0.5em{\scshape i\kern-0.25em b}\kern-0.8em\TeX}}}

\usepackage{booktabs}
\usepackage{tabularx}
\usepackage{array}
\newcolumntype{Y}{>{\raggedright\arraybackslash}X}

\copyrightyear{2026}
\acmYear{2026}
\setcopyright{cc}
\setcctype{by}
\acmConference[RecSys '26]{20th ACM Conference on Recommender Systems}{September 27-October 02, 2026}{Minneapolis, MN, USA}
\acmBooktitle{20th ACM Conference on Recommender Systems (RecSys '26), September 27-October 02, 2026, Minneapolis, MN, USA}
\acmDOI{10.1145/3773078.3841250}
\acmISBN{979-8-4007-2284-4/2026/09}
\begin{document}

\title{Whose Posts Get Ranked: Identical-Text Exposure Gaps in Bluesky Custom Feeds}

\author{Yipeng Wang}
\orcid{0009-0004-3716-8816}
\affiliation{%
  \institution{Northeastern University}
  \city{Boston}
  \state{Massachusetts}
  \country{USA}
}
\email{wang.yipen@northeastern.edu}

\author{Mohit Singhal}
\orcid{0000-0002-7423-9116}
\affiliation{%
  \institution{Northeastern University}
  \city{Boston}
  \state{Massachusetts}
  \country{USA}
}
\email{m.singhal@northeastern.edu}

\renewcommand{\shortauthors}{Yipeng Wang and Mohit Singhal}



\begin{abstract}
Bluesky lets users deploy custom feeds, independently operated recommendation algorithms that the platform serves alongside thousands of others. This paper investigates how evenly these feeds treat posts with the same text. To measure this, we take repeated snapshots of the Top-50 lists that 1,366 public feeds return, and we group posts with identical text, from different authors, that were created before the same list response and closely matched in age. Exposure diverges widely inside these matched sets, which span 250 feeds: in 33\% of sets, one copy appears on the list while another does not. Fixed-effects regressions show that this divergence is associated with the author's history on the specific feed. Authors new to a feed receive less exposure for the same text ($-0.061$ in reciprocal-rank weight), while authors whose posts the feed has returned before receive more. A new author with more followers than the competing author still loses 74\% of head-to-head comparisons. Media and post-type features show no detectable association after multiple-comparison correction. These results are early evidence that access to many independent feeds is not enough to give identical texts equal exposure.

\end{abstract}

\maketitle

\section{Introduction}

In a ranked feed, a post's rank largely determines how many users see it. Fair-ranking research treats ranking as an exposure allocation problem~\cite{singh2018fairness,biega2018equity,diaz2020expected,singhal2025auditing}, and this line of work has been systematically summarized in multiple surveys~\cite{zehlike2023fairranking1,zehlike2023fairranking2,li2023fairness}. Most existing research focuses on optimizing an ideal exposure allocation scheme for a single platform-controlled ranker; how the rankers already deployed on a platform allocate exposure is rarely measured. Bluesky is well suited to this measurement. Its custom feeds let anyone publish a feed generator, an independent recommendation service that the platform distributes and that others can subscribe to~\cite{kleppmann2024bluesky}; at present, roughly 40{,}000 feeds from 18{,}000 creators coexist on the platform~\cite{balduf2024looking}. This raises an untested question: can the same text receive equal exposure? We test this directly in 250 feeds where duplicate posts co-occur.

Holding content fixed is an established audit strategy. Correspondence audits submit identical r\'esum\'es that differ only in the applicant's name~\cite{bertrand2004emily}; in artificial music markets, identical songs can achieve vastly different levels of success~\cite{salganik2006experimental}; and the success of the same image repeatedly posted on Reddit depends on the title and timing of publication~\cite{lakkaraju2013whatsinaname}. 
Algorithm audits have extended this method to social feed scenarios~\cite{sandvig2014auditing,huszar2022algorithmic}. 
Existing Bluesky-related measurements map the platform network and feed ecosystem~\cite{balduf2024looking}, but do not measure how feeds allocate exposure. To the best of our knowledge, this is the first exposure audit study based on naturally occurring duplicate posts, without the need to create accounts or inject testing content; it is also the first study to measure within-feed exposure allocation across user-created feeds. We asked:  \\

\noindent\textbf{RQ1.} To what extent do near duplicate posts by different authors receive unequal rank-based exposure within the same Bluesky custom-feed snapshot?

\noindent\textbf{RQ2.} Are within-set exposure differences associated with authors' previous exposure on the same feed after holding text, feed, snapshot time, and approximate post age constant?

Our procedure is as follows: (1) we find identical-text posts from different authors in repeated Top-50 snapshots; (2) we form matched sets in which these posts were time-eligible for the same feed response, with creation and age spans capped at 120 minutes (60/30-minute robustness); and (3) we score each copy's rank-weighted exposure and regress the copy-level differences on author-history, profile, media, and post-type features under matched-set fixed effects.

\section{Methods}
\textbf{Data collection and feed snapshots.} We capture the ranked output of public Bluesky custom feeds roughly every 10 minutes via the \texttt{app.bsky.feed.getFeed} endpoint of the Bluesky API~\cite{bsky2026getfeed}.
A single collection from a single feed is called a snapshot, and posts that enter its Top-50 list are collected and returned. Before filtering, we observed a total of 7.3 million distinct Top-50 posts. We only keep complete snapshots, defined as snapshots containing 50 distinct ranks and 50 distinct posts after deduplication, forming 50 non-repeating rank--post pairs. After filtering, we obtained 297{,}093 complete snapshots (February--April 2026) covering 1{,}366 feeds.

\textbf{Matched sets.} A matched set is a set of posts with byte-identical text but different authors, all competing on the same complete snapshot (i.e., holding feed and snapshot time fixed). We refer to these posts within the group as copies. Each copy was published before the snapshot was generated and was no more than 7 days old; the creation-time span and age span of each matched set are capped at 120 minutes, and each group is required to include at least two different authors to ensure that each copy competes in the same ranking at a similar age. Byte-exact matching is the core matching rule.
The main sample contains a total of 64{,}853 matched sets. The vast majority of duplicate content consists of news and sports headlines, fan hashtags, and greeting posts posted across accounts, often from paired accounts publishing the same text in parallel. The most frequent single text accounts for 0.5\% of all matched sets, and the largest single feed contributes 11\% of the matched sets. Because the text is byte-identical, differences in the matched text itself cannot explain the exposure gap. There may still be differences in media attachments, thread positions, and moderation status between different copies~\cite{singhal2023sokmoderation}; we examine observable media factors below.

\textbf{Exposure outcome.} The exposure of a returned post is measured by the position weight corresponding to its rank. The calculation methods include reciprocal rank $1/r$ (RR, main indicator) and discounted cumulative gain weight $1/\log_2(r{+}1)$ (DCG). A copy that is not returned by the ranking receives zero exposure.

\textbf{Estimation.} We regress the RR exposure of each copy on the feature blocks. The model adopts matched-set fixed effects and assigns equal weight to each matched set. The fixed effects absorb all factors that are common to all copies within the same matched set: the feed, snapshot time, text content, and all other variables that have a common impact on all copies at that time point. Standard errors (SEs) are clustered by text group, which refers to the collection of all posts that share byte-identical text across all snapshots. We use the Benjamini--Hochberg procedure to control the false discovery rate (FDR), and perform corrections within feature blocks ($q_{\mathrm{blk}}$) and across all tests ($q_{\mathrm{all}}$).

\section{Results}\label{sec:results}

\begin{figure}[t]
  \centering
  \includegraphics[width=\columnwidth]{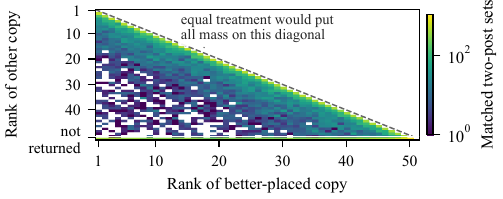}
  \caption{Ranks of identical-text copies on the same feed snapshot, for the 58{,}318 sets that have exactly two posts and a returned copy (89.9\% of all 64{,}853 sets; $\le$120-minute spans). Color: number of sets per rank pair (log scale; white = none).}
  \label{fig:heatmap}
  \Description{Heatmap of the rank of the better-placed copy versus the rank of the other copy, showing wide dispersion and a large share of copies that are not returned.}
\end{figure}

\textbf{Exposure gaps.} Exposure varies substantially across copies within the same matched set. In 33\% of matched sets, one copy is returned while another is missing from the list; in 6.8\% of matched sets, one copy reaches the Top-10 while another does not appear on the list at all. Figure~\ref{fig:heatmap} shows 58{,}318 matched sets that contain exactly two posts and at least one returned copy. The average DCG-weight gap between paired copies is 0.102 and reaches 0.25 in the most unequal decile of matched sets. A gap of 0.25 is equivalent to a post's entire DCG weight at rank~15. On average, the better-ranked copy captures 66\% of the total exposure in its matched set. When the time-span caps are tightened to 60 minutes and 30 minutes, the exposure gap barely decreases (0.095 and 0.085, respectively).

\textbf{Author history.} We find that the exposure gap is highly correlated with each author's prior history on the corresponding feed. In the fixed-effects model (Table~\ref{tab:main}), if that feed had not returned any of an author's posts in our logs before the snapshot, the author's copy receives $-0.061$~RR                       for the same text ($p=0.003$). Prior exposure on the same feed positively predicts current exposure: for each natural-log unit increase in prior exposure, current RR exposure increases by $+0.032$ ($p=0.007$, $q_{\mathrm{all}}=0.078$). Authors entering a feed for the first time are also less likely to be returned at all; under the strictest 30-minute time-span cap, the return probability decreases by 27 percentage points ($q_{\mathrm{blk}}=0.007$). These two estimates remain stable under different time-span caps, two text-matching rules, and the same-media specification, and they also pass robustness checks using permutation tests and multi-way clustering. Duplicate suppression by arrival order also cannot explain this penalty. For example, after controlling for each copy's arrival order, the penalty remains $-0.049$ ($p=0.016$). This phenomenon is not driven by a single feed. More specifically, after dropping one feed at a time, all 250 refits remain negative, with 99.6\% satisfying $p<0.05$. After excluding the author who published the most duplicate content, the estimate remains nearly unchanged ($-0.063$, $p=0.003$). Among the 6{,}768 matched sets that pair a new author with a previously returned author, the new author's copy ranks better only 16\% of the time. Even among the 746 matchups where the new author has more followers, the new author's copy still ranks lower 74\% of the time.

\textbf{Content features.} We test 304 media and post-type features (such as has\_image, is\_reply, and number of hashtags) that differ between the copies of a set; none are statistically significant.

\textbf{Limitations.} The majority of the custom feeds we collected use algorithms that are black boxes. This means there may be confounding variables that affect our results. Author history is observed only within our collection window (left-censored). 

\section{Conclusion}

Bluesky provides thousands of independently operated custom feeds that rank posts from the same pool, but whether these feeds give identical text the same exposure had not been tested before. We audited 64{,}853 matched sets of identical-text posts observed in the same batch of snapshots across 250 feeds. The results showed substantial differences in exposure across copies within the same group, and in 33\% of the groups, one copy was returned while another was missing. This gap is related to the author’s past history on the feed. Holding the feed, time, and text fixed, an author whose posts had never been returned by that feed in our logs received a $-0.061$ reciprocal-rank penalty. Among the 304 media and post-type features we observed, none could explain this gap. Within each individual feed, this phenomenon resembles the popularity-bias feedback loops observed in centralized recommendation systems.  Since this audit uses only ranked outputs and naturally occurring duplicate posts, feed operators can also apply this method to their own logs to monitor such gaps.

\begin{table}[t] \caption{Author-history fixed-effects estimates for reciprocal-rank exposure.} \label{tab:main} \centering \footnotesize \setlength{\tabcolsep}{2.5pt} \begin{tabularx}{\columnwidth}{@{}Y r c c c@{}} \toprule \textbf{FE term} & \shortstack{RR coef.\\(SE)} & $q_{\mathrm{blk}}$ & $q_{\mathrm{all}}$ & \shortstack{Perm.\\$p$} \\ \midrule New author on feed & $-0.061$ (0.020) & 0.023 & 0.047 & 0.007 \\ Prior same-feed exp.\ (log) & $+0.032$ (0.012) & 0.036 & 0.078 & 0.012 \\ \midrule \multicolumn{5}{@{}p{\columnwidth}@{}}{\scriptsize Multiway-cluster $p$ (new author): exact text 0.003; +author 0.007; +feed 0.028; text+author+feed 0.024. Family-wise max-T perm.\ $p$: 0.017.}\\ \multicolumn{5}{@{}p{\columnwidth}@{}}{\scriptsize Media and post-type blocks: 0 of 304 testable terms pass FDR.}\\ \bottomrule \end{tabularx} \end{table}

\bibliographystyle{ACM-Reference-Format}
\bibliography{references}

@inproceedings{singh2018fairness,
  author    = {Singh, Ashudeep and Joachims, Thorsten},
  title     = {Fairness of Exposure in Rankings},
  booktitle = {Proceedings of the 24th ACM SIGKDD International Conference on Knowledge Discovery \& Data Mining (KDD '18)},
  year      = {2018},
  pages     = {2219--2228},
  address   = {London, United Kingdom},
  publisher = {Association for Computing Machinery},
  doi       = {10.1145/3219819.3220088}
}

@inproceedings{singhal2025auditing,
  title={Auditing Yelp’s Business Ranking and Review Recommendation Through the Lens of Fairness},
  author={Singhal, Mohit and Pacheco, Javier and Khorzooghi, Seyyed Mohammad Sadegh Moosavi and Debi, Tanusree and Asudeh, Abolfazl and Das, Gautam and Nilizadeh, Shirin},
  booktitle={Proceedings of the International AAAI Conference on Web and Social Media},
  volume={19},
  pages={1798--1816},
  year={2025}
}

@inproceedings{biega2018equity,
  author    = {Biega, Asia J. and Gummadi, Krishna P. and Weikum, Gerhard},
  title     = {Equity of Attention: Amortizing Individual Fairness in Rankings},
  booktitle = {The 41st International ACM SIGIR Conference on Research \& Development in Information Retrieval (SIGIR '18)},
  year      = {2018},
  pages     = {405--414},
  publisher = {Association for Computing Machinery},
  doi       = {10.1145/3209978.3210063}
}

@inproceedings{diaz2020expected,
  author    = {Diaz, Fernando and Mitra, Bhaskar and Ekstrand, Michael D. and Biega, Asia J. and Carterette, Ben},
  title     = {Evaluating Stochastic Rankings with Expected Exposure},
  booktitle = {Proceedings of the 29th ACM International Conference on Information \& Knowledge Management (CIKM '20)},
  year      = {2020},
  pages     = {275--284},
  publisher = {Association for Computing Machinery},
  doi       = {10.1145/3340531.3411962}
}

@article{salganik2006experimental,
  author    = {Salganik, Matthew J. and Dodds, Peter Sheridan and Watts, Duncan J.},
  title     = {Experimental Study of Inequality and Unpredictability in an Artificial Cultural Market},
  journal   = {Science},
  volume    = {311},
  number    = {5762},
  pages     = {854--856},
  year      = {2006},
  publisher = {American Association for the Advancement of Science},
  doi       = {10.1126/science.1121066}
}

@inproceedings{lakkaraju2013whatsinaname,
  author    = {Lakkaraju, Himabindu and McAuley, Julian and Leskovec, Jure},
  title     = {What's in a Name? Understanding the Interplay between Titles, Content, and Communities in Social Media},
  booktitle = {Proceedings of the Seventh International AAAI Conference on Weblogs and Social Media (ICWSM)},
  volume    = {7},
  number    = {1},
  pages     = {311--320},
  year      = {2013},
  doi       = {10.1609/icwsm.v7i1.14408}
}

@article{bertrand2004emily,
  author    = {Bertrand, Marianne and Mullainathan, Sendhil},
  title     = {Are Emily and Greg More Employable Than Lakisha and Jamal? A Field Experiment on Labor Market Discrimination},
  journal   = {The American Economic Review},
  volume    = {94},
  number    = {4},
  pages     = {991--1013},
  year      = {2004},
  publisher = {American Economic Association},
  doi       = {10.1257/0002828042002561}
}

@inproceedings{sandvig2014auditing,
  author    = {Sandvig, Christian and Hamilton, Kevin and Karahalios, Karrie and Langbort, Cedric},
  title     = {Auditing Algorithms: Research Methods for Detecting Discrimination on Internet Platforms},
  booktitle = {Data and Discrimination: Converting Critical Concerns into Productive Inquiry, a preconference at the 64th Annual Meeting of the International Communication Association (ICA)},
  year      = {2014},
  address   = {Seattle, WA, USA}
}

@article{huszar2022algorithmic,
  author  = {Husz{\'a}r, Ferenc and Ktena, Sofia Ira and O'Brien, Conor and Belli, Luca and Schlaikjer, Andrew and Hardt, Moritz},
  title   = {Algorithmic amplification of politics on {Twitter}},
  journal = {Proceedings of the National Academy of Sciences},
  volume  = {119},
  number  = {1},
  pages   = {e2025334119},
  year    = {2022},
  doi     = {10.1073/pnas.2025334119}
}

@inproceedings{kleppmann2024bluesky,
  author    = {Kleppmann, Martin and Frazee, Paul and Gold, Jake and Graber, Jay and Holmgren, Daniel and Ivy, Devin and Johnson, Jeromy and Newbold, Bryan and Volpert, Jaz},
  title     = {Bluesky and the {AT} {Protocol}: Usable Decentralized Social Media},
  booktitle = {Proceedings of the ACM CoNEXT 2024 Workshop on the Decentralization of the Internet (DIN '24)},
  pages     = {1--7},
  year      = {2024},
  publisher = {ACM},
  address   = {Los Angeles, CA, USA},
  doi       = {10.1145/3694809.3700740}
}

@inproceedings{balduf2024looking,
  author    = {Balduf, Leonhard and Sokoto, Saidu and Ascigil, Onur and Tyson, Gareth and Scheuermann, Bj{\"o}rn and Korczy{\'n}ski, Maciej and Castro, Ignacio and Kr{\'o}l, Micha{\l}},
  title     = {Looking {AT} the Blue Skies of Bluesky},
  booktitle = {Proceedings of the 2024 ACM on Internet Measurement Conference (IMC '24)},
  pages     = {76--91},
  year      = {2024},
  publisher = {ACM},
  address   = {Madrid, Spain},
  doi       = {10.1145/3646547.3688407}
}

@article{li2023fairness,
  author    = {Li, Yunqi and Chen, Hanxiong and Xu, Shuyuan and Ge, Yingqiang and Tan, Juntao and Liu, Shuchang and Zhang, Yongfeng},
  title     = {Fairness in Recommendation: Foundations, Methods, and Applications},
  journal   = {ACM Transactions on Intelligent Systems and Technology},
  volume    = {14},
  number    = {5},
  articleno = {95},
  year      = {2023},
  publisher = {Association for Computing Machinery},
  doi       = {10.1145/3610302}
}

@article{zehlike2023fairranking1,
  author    = {Zehlike, Meike and Yang, Ke and Stoyanovich, Julia},
  title     = {Fairness in Ranking, Part {I}: Score-Based Ranking},
  journal   = {ACM Computing Surveys},
  volume    = {55},
  number    = {6},
  articleno = {118},
  year      = {2023},
  publisher = {Association for Computing Machinery},
  doi       = {10.1145/3533379}
}

@article{zehlike2023fairranking2,
  author    = {Zehlike, Meike and Yang, Ke and Stoyanovich, Julia},
  title     = {Fairness in Ranking, Part {II}: Learning-to-Rank and Recommender Systems},
  journal   = {ACM Computing Surveys},
  volume    = {55},
  number    = {6},
  articleno = {117},
  year      = {2023},
  doi       = {10.1145/3533380}
}

@misc{bsky2026getfeed,
  author       = {{Bluesky Social, PBC}},
  title        = {app.bsky.feed.getFeed --- {Bluesky} {HTTP} {API} Reference},
  year         = {2026},
  howpublished = {\url{https://docs.bsky.app/docs/api/app-bsky-feed-get-feed}},
  note         = {Accessed July 5, 2026}
}

@inproceedings{singhal2023sokmoderation,
  author    = {Singhal, Mohit and Ling, Chu and Paudel, Aayush and Thota, Rajath and Kumarswamy, Shashank and Stringhini, Gianluca and Nilizadeh, Shirin},
  title     = {{SoK}: Content Moderation in Social Media},
  booktitle = {2023 {IEEE} European Symposium on Security and Privacy (EuroS\&P)},
  pages     = {868--885},
  publisher = {{IEEE}},
  year      = {2023},
  doi       = {10.1109/EuroSP57164.2023.00056}
}

\end{document}